\documentclass[12pt,a4paper]{article}

\usepackage{graphicx}
\usepackage{amsmath,amssymb,mathrsfs}
\usepackage{cite}
\usepackage{bm}
\usepackage{color}
\usepackage{hyperref}
\usepackage{enumerate}
\usepackage{enumitem}
\usepackage{verbatim}
\usepackage{gensymb}

\usepackage{setspace}
\date{\today}

\def\be{\begin{equation}}
\def\ee{\end{equation}}

\DeclareSymbolFont{matha}{OML}{txmi}{m}{it}

\DeclareMathSymbol{v}{\mathord}{matha}{118}

\author{Aron C. Wall\footnote{aroncwall@gmail.com}
\\ \textit{Stanford Institute for Theoretical Physics}
\\ \textit{382 Via Pueblo, Stanford University, Stanford, CA, 94305} }
\title{A Survey of Black Hole Thermodynamics}

\begin{document}

\maketitle

\begin{abstract}
This is an introductory, up-to-date review of the essentials of black hole thermodynamics.  The main topics surveyed are: (i) the four laws of thermodynamics as applied to a black hole horizon, and the current status of their proofs; (ii) different definitions of horizons, and their unique properties; (iii) the nature of black hole entropy, its quantum and stringy corrections, and ultimate origin from quantum gravity microstates; (iv) the focusing law for the area/entropy; and finally (v) the holographic principle, and how we can use it to learn about the information inside black holes.
\end{abstract}

\begin{center}
\medskip\medskip\medskip
\centering{\it Dedicated to the memory of Stephen Hawking}
\end{center}

\newpage
\tableofcontents

\section{Introduction}

Black holes are peculiar entities, but in one respect they are strangely normal: they obey laws of thermodynamics similar to ordinary matter systems \cite{bardeen1973four}, when viewed from the perspective of an observer outside the horizon, so long as we attribute to the horizon an entropy $S$ proportional to its area $A$, a temperature $T$ proportional to its surface gravity $\kappa$, and of course an energy $E$ proportional to its mass $M$.

This came as a surprise for several reasons.  First, the region outside the horizon is seemingly an open system (since matter can fall in), whereas the second law of thermodynamics normally applies only to closed systems.  Second, at the classical level stationary black holes have only a few degrees of freedom (e.g. mass and charge) and so it is mysterious what statistical mechanical states are counted by this entropy.

Recently a great deal of progress has been made in understanding the precise way in which black hole thermodynamics should be interpreted, especially in dynamical situations where matter may be falling across the horizon.  Some of these insights have come from a better understanding of classical general relativity (GR), and also how quantum field theory (QFT) works on a black hole background.  Even more have come from the \emph{holographic principle}, the idea that the degrees of freedom in a gravitational system are somehow encoded in the spatial boundary of the system.

For simplicity this review is centered around black holes, but it should be remembered that most of the results can be generalized to other contexts, e.g. to cosmological horizons, or to noncompact surfaces in asymptotically flat or anti-de Sitter (AdS) spacetimes. An excellent older review is Jacobson \cite{Jacobson}, but below we will describe several more recent developments, especially for non-stationary horizons.

In Section \ref{sec:energy}, we briefly introduce some energy conditions that will be repeatedly used in the proofs of theorems about black holes. Section \ref{sec:thermal} describes the nature of thermal equilibrium for black holes. Section \ref{sec:1st} introduces the first law of black hole thermodynamics, which allows the comparison of nearby equilibrium solutions. In Section \ref{sec:BHent}, we discuss the black hole entropy and its corrections in quantum and/or stringy situations. Section \ref{sec:2nd} describes dynamical black holes, different definitions of horizons, and the second law in various regimes. A key concept is the focusing of entropy along lightlike hypersurfaces.  Finally in Section \ref{sec:holographic} we discuss the holographic principle and what it teaches us about black holes in quantum gravity.

%
%

\section{Energy Conditions} \label{sec:energy}

When proving theorems about general relativity, it is often necessary to assume some positivity conditions on the stress-energy tensor $T_{ab}$;\footnote{Some proofs reviewed below contain additional technical assumptions; see the references for details.} otherwise all possible metrics are possible solutions to the Einstein equations:
\be
R_{ab} - \frac{1}{2}g_{ab}R = 8\pi G\,T_{ab}.
\ee  
The following energy conditions tend to be obeyed by most reasonable classical fields: 

\paragraph{Null Energy Condition (NEC):} $T_{ab} \,\hat{v}^a \hat{v}^b \geq 0$ for all null (lightlike) vectors $\hat{v}^a$.\,\footnote{This is a vee ($v$), not a nu ($\nu$).  The default LaTeX italicized \textit{v} looks too much like a \textit{u}.}



\paragraph{Dominant Energy Condition (DEC):} $T_{ab} \,\hat{t}_1^a \,\hat{t}_2^b \geq 0$ for all timelike vectors $\hat{t}_1$, $\hat{t}_2$. \newline

\noindent Yet both of these energy conditions can be violated by quantum fields \cite{epstein1965nonpositivity}, and also by classical fields that are non-minimally coupled to curvature \cite{flanagan1996does}.\footnote{The black hole entropy receives corrections in these cases, as will be discussed in section \ref{sec:BHent}.}  However, in QFT, the integral of the null energy is still positive on any infinitely-extended null geodesic  \cite{borde1987geodesic}; this is known as the \emph{Averaged Null Energy Condition} (ANEC):\,\footnote{For QFT in curved spacetime, the ANEC holds if the null geodesic is achronal, i.e. having no timelike separated points \cite{graham2007achronal,Kontou:2015yha}.}
\be\label{ANEC}
\int_{-\infty}^{\infty} T_{ab} \,\hat{v}^a \hat{v}^b\,dv \ge 0,
\ee
where $v$ is an uniform (i.e. affine) coordinate labelling the null direction, and $\hat{v}^a$ is the corresponding unit vector.\footnote{For a long time the ANEC was only known to hold in special cases (like free field theory), but it has recently been proven for general quantum field theories \cite{Tom1,Tom2}.}  In a gravitational theory, the ANEC can be used to rule out warp drives and other causality violations \cite{Gao:2000ga}.

\section{Thermal Equilibrium} \label{sec:thermal}

In this section we describe thermal equilibrium for classical and quantum black holes.

\subsection{Killing Horizons} \label{sec:killing}

\begin{figure}[ht]
\centering
\includegraphics[width=.8\textwidth]{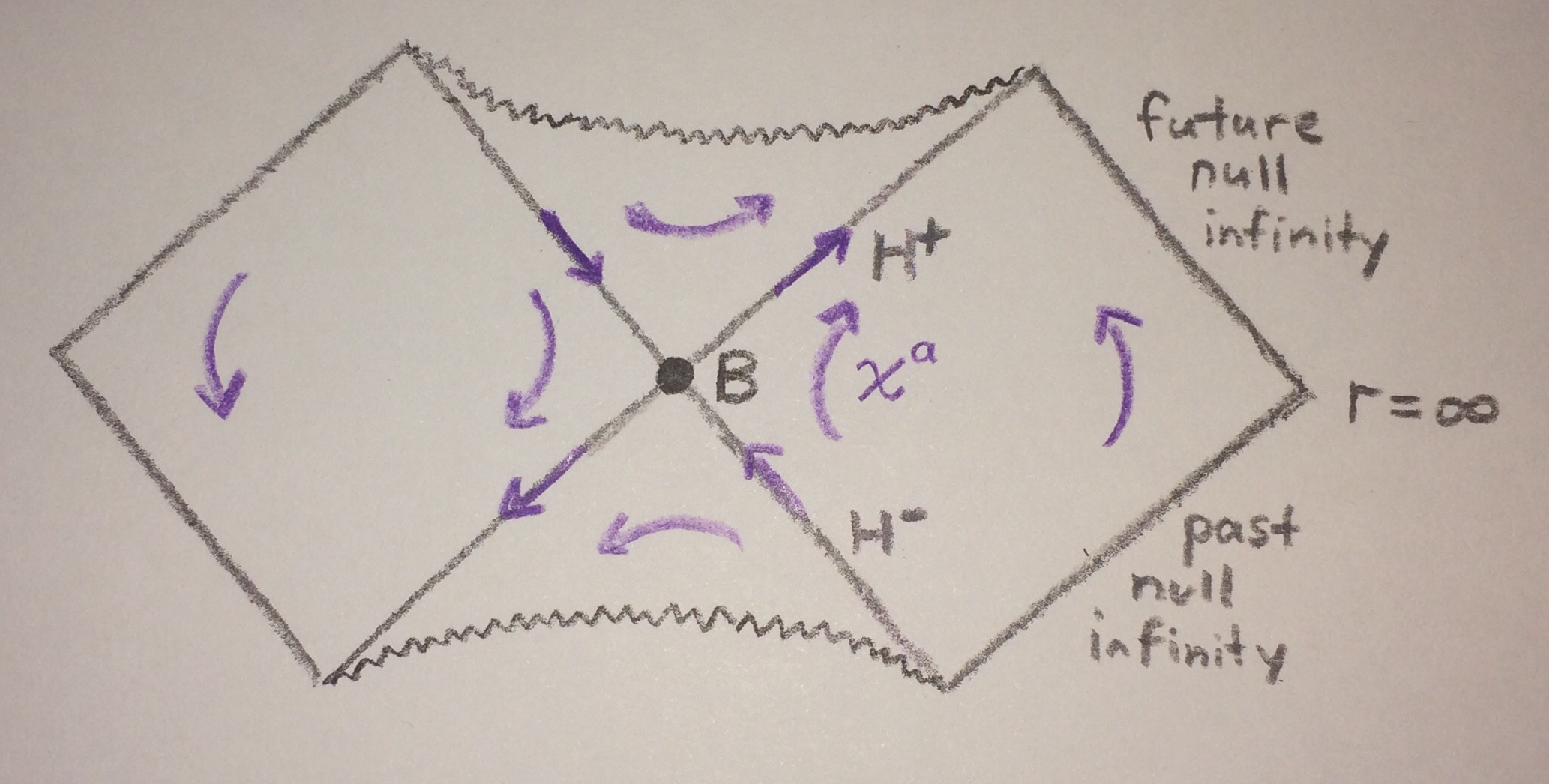}
\caption{\small Penrose diagram of the Schwarzschild black hole, an example of a Killing horizon.  Each point represents a sphere, light travels at $45\degree$, solid boundary edges are infinity, and jagged edges are singularities.  The future horizon $H^+$, past horizon $H^-$, bifurcation surface $B$, and the action of the Killing vector $\chi^a$ are shown.  (In the case of a black hole that forms from collapse and later becomes stationary, only the right and upper quadrants exist.)}\label{fig:Killing}
\end{figure}

The event horizon of a black hole in $D$ spacetime dimensions is a $(D-1)$-dimensional surface, composed of the null geodesics which just barely fail to escape from the black hole.  When the spacetime has a symmetry that maps the horizon into itself along the null direction, we call it a \emph{Killing horizon}.  In an eternal (maximally extended, non-extremal) black hole spacetime, the Killing horizon consists of a future horizon $H^+$, a past horizon $H^-$, and a (D-2)-dimensional bifurcation surface $B$ where the two intersect.  See Fig. \ref{fig:Killing}.

The Killing symmetry of the horizon is generated by a \emph{Killing vector} $\chi^a$ \footnote{A Killing vector satisfies the equation $\nabla_a \chi_b + \nabla_b \chi_a = 0$, which implies that it generates a symmetry of the metric.}.  At infinity, $\chi^a = \hat{t}^a + \Omega \hat{\phi}^a$ is a unit time translation, plus a rotation if the black hole has nonzero angular velocity $\Omega$.  But at the bifurcation surface $B$, it always looks locally like a Lorentz boost in the plane normal to $B$.  
 

In black hole thermodynamics, the temperature $T$ of a Killing horizon is identified with the surface gravity $\kappa$ evaluated on $H^{\pm}$:
\be
\frac{2\pi T}{\hbar} = \kappa \equiv \left\lvert \nabla_a \chi_b \right\rvert \equiv \sqrt{-\frac{1}{2} (\nabla_a \chi_b)(\nabla^a \chi^b)}.
\ee
For a Schwarzschild black hole of radius $R$, $\kappa = 1/2R$.

The \emph{zeroth law} of classical black hole thermodynamics states that this surface gravity is constant everywhere on $H^\pm$.  This law can be proven for any Killing horizon which is either i) static or ii) axisymmetric with a $t \to -t$, $\phi \to -\phi$ reflection symmetry of the time and angular coordinates \cite{carter1971axisymmetric,racz1996global}.  There is a more general proof which holds whenever the horizon is stationary (regardless of whether the spacetime outside the horizon admits a Killing symmetry), but it uses the DEC.\footnote{It is not clear how to extend this result beyond Einstein gravity, cf. \cite{sarkar2013issue} for a negative result in Lovelock gravity.}

\subsection{Hartle-Hawking State}\label{sec:HH}

At the quantum level, black holes radiate Hawking quanta \cite{hawking1975particle}.  These quanta emerge from the short-distance modes near the horizon, which are red-shifted as they escape from the horizon to infinity.

A static Killing horizon admits a special \emph{Hartle-Hawking (HH) state}, where the quantum fields are in thermal equilibrium with the black hole with a temperature $T = \hbar \kappa / 2 \pi$.\footnote{Here and below, we set the speed of light $c$ and Boltzman's constant $k_B$ to 1.}  This state can be obtained by Wick rotating the black hole geometry, and then doing a path integral on the resulting Euclidean signature geometry \cite{hartle1976path,israel1976thermo}.  The thermality of the state $\rho_{HH}$ outside of the bifurcation surface $B$ is then guaranteed by the periodicity of the geometry in the direction of imaginary Killing time.\footnote{This statement assumes exact Lorentz invariance, without which the laws of black hole thermodynamics can be violated \cite{Dubovsky:2006vk,Eling:2007qd,Jacobson:2008yc}.}  That is,
\be\label{thermal}
\rho_{HH} \propto e^{-K / T},
\ee
where $K$ is proportional to the Killing energy:
\be\label{KillingK}
K = \int_\Sigma T_{ab}\,\chi^a\,d\Sigma^b,
\ee
$\Sigma$ is any partial time slice connecting $B$ to spatial infinity, $d\Sigma^a$ is the natural volume measure for fluxes across $\Sigma$, and $\chi^a$ is the static Killing vector.

While the HH-state is thermal from the perspective of an observer restricted to the region outside the bifurcation surface, it is actually the ground state with respect to translations $v \to v + c$ of a uniform null coordinate along $H^+$  \cite{israel1976thermo,fulling1977alternative,sewell,kay1991theorems}.  This actually implies that in the HH-state, the ANEC integral \eqref{ANEC} is exactly zero on \emph{each} individual lightray $\gamma$ of the horizon---the LHS of \eqref{ANEC} is just the energy associated with the null translation symmetry along $\gamma$.\footnote{This is a symmetry of the horizon $H^{+}$, but not the rest of the spacetime.  It is therefore not associated with a globally conserved quantity.}

\begin{figure}[ht]
\centering
\includegraphics[width=.6\textwidth]{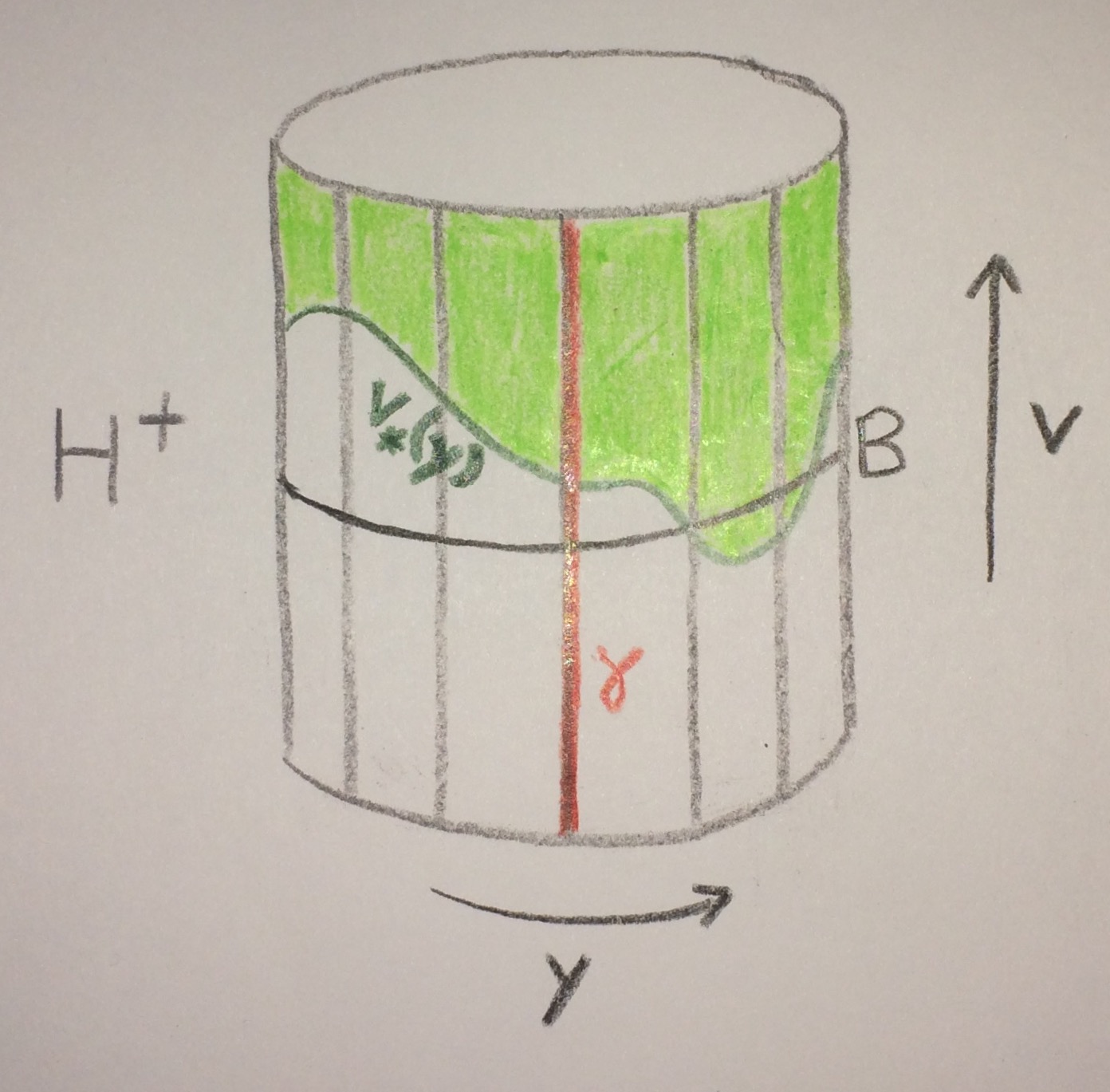}
\caption{\small The lightrays generating the future Killing horizon $H^+$ are shown (vertical lines) with $(v,y)$ coordinates.  $\rho_{HH}$ is a ground state with respect to translation along any lightray $\gamma$, but is thermal with respect to a ``boost'' in the (shaded) region above any cut $v_*(y)$.}\label{fig:raycut}
\end{figure}

We can now ask what happens if we cut the horizon at an arbitrary cut $v = v_*(y)$, where $y$ is the $(D-2)$ transverse coordinates labelling different lightrays (Fig. \ref{fig:raycut}).  In the region outside any such cut, it can be shown that $\rho_{HH}$ is thermal with respect to an integral of the stress tensor on $H^+$ above the cut $v_*(y)$, plus a piece $K_\infty$ associated with energy that goes off to future infinity without ever crossing the horizon:
\be\label{K}
K(v_*) = 2\pi \int_{v > v_*(y) \text{ on } H^+}\!\!\!\!\!\!\! \!\!\!\!\!\!\! \!\!\!\!\!\!\! \!\!\!
T_{ab}\,\widetilde{\chi}^a\,d\Sigma^b \:\:+\:\: K_\infty.
\ee
Here $\widetilde{\chi}^a \equiv (v - v_*(y))\hat{v}^a$ ($\hat{v}^a$ being the unit $v$-vector) is the vector generating an approximate near-horizon symmetry that looks like a Lorentz boost around $v_*(y)$---it is a linear combination of the boost Killing vector $\chi^a = v\,\hat{v}^a$ and the null translation symmetries of $H^+$ mentioned in the previous paragraph, which take the form $f(y) \hat{v}^a$.  The natural null integration measure on $H^+$ is just $d\Sigma^b = \hat{v}^b\,dv\,dA$.  Note that \eqref{K} reduces to \eqref{KillingK} when the cut $v_*(y)$ is chosen to be the bifurcation surface $B$.\footnote{Eq. \eqref{K} was derived in \cite{wall2011} if the QFT is free or superrenormalizable, and more recently by \cite{Casini:2017roe} for all field theories flowing to a conformal fixed point in the ultraviolet. (Technically \cite{Casini:2017roe} only considered the case of Rindler horizons in Minkoswki, where all information falls across the horizon and hence there is no constant term $K_\infty$, but I expect their results can be generalized to the black hole case.  Their work also implies a novel entanglement entropy proof of the $a$-theorem in $D = 4$ dimensions \cite{Casini:2017vbe}.)}

\section{First Law and Canonical Energy}\label{sec:1st}
The \emph{first law} of black hole mechanics (really the Clausius relation) governs first order variations to thermal equilibrium states. In the case of a black hole of mass $M$, charge $Q$ and angular momentum $J$, it takes the form
\be \label{eq:clausius}
dM = T \,dS + \Omega \,dJ + \Phi \, dQ,
\ee
where $\Omega$ is the angular velocity of light rays on the horizon, $\Phi$ is the electric potential on the horizon, and $T$ and $S$ are proportional to the surface gravity $\kappa$ and area $A$ respectively.  It has been proved that the first law holds both for nearby stationary solutions (the \emph{stationary comparison} first law) \cite{bardeen1973four}, and, more interestingly, for matter fields dynamically falling across the horizon (the \emph{physical process} first law) assuming that the black hole begins and ends in a stationary configuration \cite{wald1994quantum,Gao:2001ut}.  If we are only interested in the physical process law near the horizon, we can more conveniently write it as $dK = T dS$ where $dK = dM - \Omega dJ - \Phi dQ$ is the Killing energy flux given by an integral of the stress tensor on the horizon \eqref{KillingK}.

The first law may be proven most elegantly using Noetherian methods \cite{wald1993black,JKM,iyer1994some,Gao:2001ut}.  Consider the first order variation $\delta H$ of the ADM canonical energy associated with the black hole's Killing vector $\chi_a$.  Diffeomorphism invariance implies that $\delta H$ is a total derivative, and hence can be written as a boundary integral.  The boundary value at infinity is related to the gravitational mass, but the boundary value at the horizon is the entropy $S$.  This tells us that there is a kind of gravitational ``Gauss law'' relating $\Delta \delta S \equiv \delta S_\text{late} - S_\text{early}$ to the contribution to $\delta H$ on the horizon between the two times.  If we take the ``early'' time to be the bifurcation surface $B$ and the late time to be $v = +\infty$, we recover the physical-process version of the first law \eqref{eq:clausius}.  Moreover this framework can be generalized to arbitrary theories of gravity, and one finds that for Killing horizons $S$ is given by the Wald entropy (the first term of \eqref{eq:SBH} in Section \ref{sec:stringcorrect}).


There is also a canonical energy $\cal E$ defined at second order in the perturbation, which provides a useful diagnostic criterion for the stability of classical black holes \cite{Hollands:2012sf}. At this order it can also be used to show the impossibility \cite{sorce} of overspinning or overcharging a black hole beyond extremality (the maximum allowed $Q$ and/or $J$ value for a given $M$)---even in situations where a first-order analysis seems to indicate that this is possible \cite{Hubeny:1998ga,Jacobson:2009kt}---once self-force effects are taken into consideration\cite{Zimmerman:2012zu,Barausse:2010ka,Colleoni:2015ena}.  This vindicates the \emph{third law} of black holes, which states that there is no physically allowed way to create an extremal black hole by any finite process \cite{bardeen1973four,israel}.\footnote{An alternative formulation of the third law of thermodynamics states that $S \to 0$ as $T \to 0$, but this supposed ``law'' is invalid not just for extremal black holes, but also for ordinary thermodynamic systems with ground state degeneracy.}

\section{Black Hole Entropy}\label{sec:BHent} 

The entropy of a black hole may be calculated classically by consistency with the first law (Sec. \ref{sec:1st}) or second law (Sec. \ref{sec:2nd}).  It can also be calculated by path integral methods \cite{gibbons1977action, lewkowycz}.  In this section we give the formula for black hole entropy in the classical regime, and then describe its quantum/stringy corrections, and its invariance under renormalization.  At the end we briefly discuss the corresponding microstate description in quantum gravity.

\subsection{Bekenstein-Hawking Classical Entropy}

The leading classical contribution to the entropy of a black hole in Einstein's general relativity is given by the Bekenstein-Hawking entropy, which equals one-quarter of the area of a $(D-2)$-dimensional slice of the horizon in Planck units:
\be
S_{BH} = \frac{A}{4G \hbar}.
\ee
However, when the theory of gravity is modified, or in quantum settings, there are additional correction terms to this entropy formula, which we shall now discuss:
\subsection{Quantum/Thermal Corrections} \label{sec:quantumentropy}
If there is any matter outside of the black hole, for example a neutron star orbiting it, then the entropy of the universe should include that entropy as well. Thus the total entropy is given by the \textit{generalized entropy}
\be\label{genS}
S_{\text{gen}} = \frac{\langle A \rangle}{4G \hbar} + S_{\text{out}},
\ee
where $S_{\text{out}}$ is the von Neumann entropy $-\mathrm{tr}\left(\rho \ln \rho \right)$ of the density matrix $\rho$ of the matter outside the horizon, and the area $A$ is now an operator depending on the gravitational backreaction of the quantum fields (hence the expectation value).

To properly define $S_{\text{out}}$, it is necessary to include all of the quantum field excitations outside of the horizon. These quantum excitations are responsible for Hawking radiation, and therefore must be included in a consistent analysis. But notoriously the entanglement entropy across a sharp boundary is ultraviolet divergent in QFT, even in the vacuum state. So na\"{i}vely, $S_{\text{out}} = + \infty$, due to the ``thermal atmosphere'' of modes just outside the horizon. In many calculations, when the quantum backreaction on the geometry is small, we can sidestep this issue by only considering \emph{differences} of entropy $\Delta S_{\text{out}}$ between different states. But in general we must do the same thing that we do with other divergent quantities in QFT, which is to introduce a momentum cutoff $\Lambda$ to regulate the divergences, and then renormalize them by absorbing the $\Lambda$-dependence into counterterms.

Since the leading order ($\Lambda^{D-2}$) divergence in the entanglement entropy scales with the area $A$ of the boundary, we can absorb it into a shift in the value of $1/G$ in $S_{BH}$. In spacetime dimensions $D \geq 4$, there are also subleading divergences; these are absorbed into the higher curvature corrections described in the next section.\footnote{For numerous references, see the appendix of \cite{bousso2016quantum}.  Some apparent discrepancies in this renormalization procedure have now been resolved.  For a non-minimally-coupled scalar field (with a $\xi \phi^2 R$ term in the action), consistency requires us to include e.g. the renormalization of $1/G$ arising from the Wald entropy term which is proportional to the integral of $-\xi \langle \phi^2 \rangle$ on the horizon slice \cite{ford2001classical}.  For gauge fields, it is necessary to take into account ``edge mode'' degrees of freedom living on the horizon slice \cite{casini2014remarks,donnelly2014entanglement,donnelly2015entanglement,donnelly2016geometric,jafferis2016relative} which contribute nontrivially to $S_\mathrm{out}$ and its divergences.  The graviton field ($s = 2$) should also have these edge modes \cite{jafferis2016relative}, but no fully satisfactory treatment yet exists, primarily due to the fact that the linearized graviton QFT does not make sense off-shell.}

\subsection{Higher Curvature/Stringy Corrections} \label{sec:stringcorrect}
Let us consider a gravitational action which contains, in addition to the Einstein-Hilbert term, some higher curvature corrections:
\be
I = \int d^D x \, \sqrt{-g} \, \mathcal{L}, \qquad \mathcal{L} = \left[ \frac{R}{16 \pi G} + f(R_{abcd}) \right].
\ee
These higher-curvature terms arise in classical string theory, 
as well as from loop corrections in quantum gravity.
 In an effective field theory, they might also appear as a bare term in the action.\footnote{However, certain terms in $\mathcal{L}$ classically lead to faster-than-light propagation of gravitons, if their coefficients (relative to Einstein-Hilbert) are too large compared to the string scale (defined as the lowest energy scale at which fields with spin $s>2$ appear)\cite{camanho2016causality,Papallo:2015rna}.  There might not even be a well-posed initial data problem \cite{Papallo:2017qvl,Reall:2014pwa,Reall:2014sla}.  Presumably such theories cannot be completed into a consistent non-perturbative theory of quantum gravity.}

Whatever the origin of the higher-curvature terms, they give rise to corrections in the black hole entropy formula. Using null coordinates $(u,v,y_i)$ where $g_{uv} = -1$ and $i,j \ldots$ point in the $(D-2)$ transverse directions, and defining $K_{ij}^{(u)} \equiv \tfrac{1}{2} \partial_u g_{ij}$ as the extrinsic curvature in the $u$-direction (and similar for $K_{ij}^{v}$), we use the following ``generalized area'' functional in place of $S_{BH}$:
\be \label{eq:SBH}
A_\mathrm{gen} = - \frac{2 \pi}{\hbar} \int d^{D-2} x \,\sqrt{g}\,\left[4 \frac{\partial \mathcal{L}}{\partial R_{uvuv}} + 16 \frac{\partial^2 \mathcal{L}}{\partial R_{uiuj} \partial R_{vkvl}} K_{ij}^{(u)} K_{kl}^{(v)} + \mathcal{O}(K^4)\right].
\ee
The first term is the Wald entropy \cite{wald1993black,JKM,Visser:1993nu,iyer1994some}, and is valid for stationary black holes. The expression for dynamical settings was derived for $f(\mathrm{Riemann})$ gravity in \cite{dong,Wall:2015raa}\footnote{For the special case of quadratic curvature gravity, see \cite{solodukhin2008entanglement,fursaev2013distributional,camps2014generalized}.  For discussion of the ambiguities affecting the $K^4$ order, and for some partial results when $\mathcal{L}$ includes derivatives of the Riemann tensor, see\cite{miao2015holographic,miao2015universal}.}

\subsection{Black Hole Microstates and Induced Gravity}\label{sec:induced}

The analogy to statistical mechanics suggests that black hole entropy should have a state counting description involving some quantum gravity degrees of freedom near the horizon.  Specific state-counting interpretations of black hole microstates have been proposed in both string theory \cite{Strominger:1996sh,Breckenridge:1996sn,Horowitz:1996nw,Sen:2007qy} and loop quantum gravity \cite{Ashtekar:1997yu,Meissner:2004ju,Bianchi:2012ui} (but see \cite{Jacobson:2007uj}).

We have seen that the generalized entropy $S_\mathrm{gen} = A_\mathrm{gen}(\Lambda) + S_\mathrm{out}(\Lambda)$ is independent of the renormalization cutoff scale $\Lambda$, but changing $\Lambda$ shifts entropy between the two terms.  It is therefore plausible that, if we take the cutoff $\Lambda$ to the Planck scale (the shortest possible distance that can be defined in a quantum gravity theory) then $A_\mathrm{gen} = 0$, and so the black hole entropy comes entirely from the entanglement entropy contribution to $S_\mathrm{out}$, which could be rendered finite by quantum gravity effects \cite{Sorkin:2014kta}.  Since $1/G$ and all the other terms in $\mathcal{L}$ that contribute to $A_\mathrm{gen}$ would have to vanish, this is equivalent \cite{susskind1994black,jacobson1994black,Frolov:1996aj} to the induced gravity scenario of Sakharov \cite{sakharov1968vacuum}, in which the gravitational action arises entirely from loop corrections.

\section{Dynamics and the Second Law}\label{sec:2nd}

This section describes various kinds of thermodynamic inequalities associated with entropy production for dynamical black holes, in both classical and quantum settings.

\subsection{Classical Horizons}

\begin{figure}[ht]
\centering
\includegraphics[width=.8\textwidth]{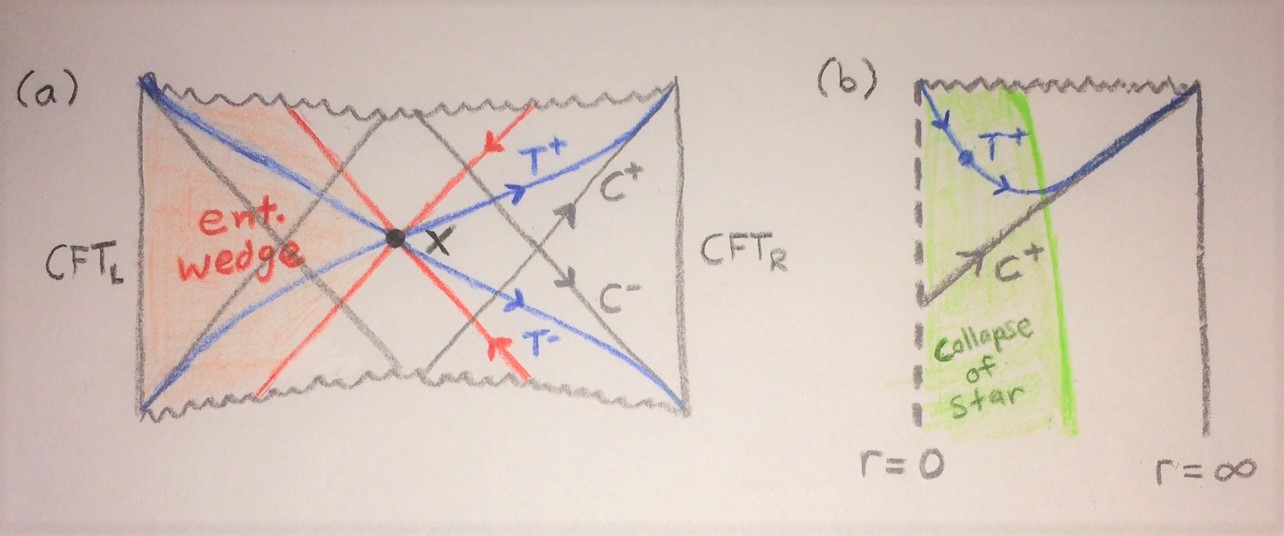}
\caption{\small (a) The different kinds of horizons $C^\pm$, $T^\pm$ are plotted on the Penrose diagram of a wormhole going between two asymptotically AdS regions (with timelike boundaries).  This is not a Killing spacetime due to the gravitational effects of matter (not shown), which tends to make the diagram wider.  Also shown is the HRT extremal surface $X$ discussed in section \ref{HEF}, and the contracting null surfaces coming out from it.  Arrows are drawn in the direction of increasing area.  In holography, this spacetime is dual to two entangled CFT's on the left and right.  The entanglement wedge dual to $\text{CFT}_L$ is shaded.  
(b) A black hole that forms from the collapse of a star.  There is neither an HRT surface, nor past horizons.  $C^+$ is always null, but $T^+$ may have timelike and spacelike segments (here a dot marks the transition).  On the top-right corner of each diagram, after the black hole settles into a stationary state, $C^+$ and $T^+$ both coincide with the late-time Killing horizon.}
\label{fig:horizons}
\end{figure}

When matter falls across a black hole horizon, it dynamically evolves, and hence the notion of a Killing horizon is no longer available.  In its place there are multiple kinds of horizons obeying a second law, shown in Fig. \ref{fig:horizons}.

First we have a future causal horizon $C^+$, defined as the boundary of the past of some locus of points at infinity \cite{jacobson2003horizon,wall2013generalized}.\footnote{This defines a \emph{future} causal horizon; we can analogously construct a past horizon $C^-$ by time-reversing the definition. The same applies to the definition of the future apparent horizon below.} In the special case where we take the locus to be all of future null infinity, this gives us the \emph{event horizon}.

If we consider an arbitrary null surface $N$ in the spacetime, labelled by an affine null coordinate $v$, then its \emph{expansion} is given by
\be
\theta \equiv \frac{1}{\delta A} \frac{d\,\delta A}{dv},
\ee  
where $\delta A$ is the area of an infinitesimal patch of $N$. The rate of change of the expansion is given by the Raychaudhuri equation:
\be \label{eq:raychau}
\frac{d \theta}{dv} = - \frac{\theta^2}{D-2} - \sigma_{ij} \sigma^{ij} - R_{vv}
\ee
where $\sigma_{ij}$ is the shear of $N$ (which measures the gravitational radiation across $N$). When the NEC is satisfied, $R_{vv} = 8 \pi G\,T_{vv} \ge 0$, so the whole RHS is negative, 
\be\label{Cfocus}
\frac{d \theta}{dv} \le 0
\ee
i.e. gravity always focuses lightrays. If $\theta < 0$ intially, then solving \eqref{eq:raychau} requires that $\theta \to - \infty$ at some finite $v$; hence these lightrays must either intersect or hit a singularity in finite time.  Since $C^+$ is defined using a future boundary condition, it turns out that its lightrays cannot intersect in the future direction, so $\theta \ge 0$. This is the \emph{classical second law} for causal horizons \cite{hawking1971gravitational}.

Now consider a compact $(D-2)$-dimensional surface $s$ in a spatially noncompact spacetime (which is reasonable for a black hole solution). If all lightrays shot out from $s$ in the future-outwards direction have $\theta < 0$, we call $s$ a \emph{trapped} surface. In this case the NEC implies the existence of a singularity to the null future of $s$ given that the spacetime has good causal properties. This is the Penrose singularity theorem \cite{penrose1965gravitational}; for a historical review see \cite{senovilla20151965}. Also, trapped surfaces always lie inside the event horizon \cite{hawking1973large,wald2010general}. If $\theta=0$ everywhere on $s$, we call $s$ a \emph{marginally trapped} surface.

We can now define a future trapping horizon $T^+$ \cite{hayward1994general} (a.k.a. a holographic screen \cite{bousso1999holography}), as a $(D-1)$-dimensional surface foliated by leaves, each of which is a future marginally trapped surface.\footnote{In the special case where the expansion $(1/\delta\! A)\, d\delta\! A/du$ in the other null direction $u$ is negative, it is also called a dynamical horizon \cite{ashtekar2002dynamical,ashtekar2003dynamical}.} Because $T^+$ is  generically not null, the fact that $\theta = 0$ does not imply the area of $T^+$ is constant in time.  Note that we can identify many possible distinct trapping horizons on the same black hole background, by e.g. taking an arbitrary foliation of the spacetime into Cauchy slices, and identifying the outermost marginally trapped surface on each one.\footnote{It is also possible to generalize the notion of surface gravity $\kappa$ to trapping horizons, allowing a local form of the physical process first law \eqref{eq:clausius}, even far from equilibrium \cite{hayward1994general,ashtekar2002dynamical}.}

The NEC also implies that trapping horizons obey various area-increase theorems; in particular a spacelike $T^+$ has increasing area when moving outwards, while a timelike $T^+$ has (oddly) increasing area when moving to the past \cite{hayward1994general}.  A future holographic screen may include segments of both signature, but the NEC implies they are always stitched together in such a way that the area increases in a consistent direction \cite{bousso2015new}.  Note that if the holographic screen settles down to a stationary Killing horizon $H^+$ at late times, this direction agrees with the time direction of the corresponding causal horizon $C^+$. The fact that the trapping horizon $T^+$ continues to exist at late times can also be used to derive the third law \cite{israel}. 

If the gravitational action includes higher-curvature corrections, the entropy is given by \eqref{eq:SBH} rather than the area. So long as we restrict attention to linearized perturbations to a Killing horizon with a non-singular bifurcation surface $B$, there always exists a focusing relation for the entropy, and hence a second law \cite{Wall:2015raa, Sarkar:2013swa}.  But in the nonlinear regime, there is probably no second law for a general gravity action.\footnote{In certain special cases, such as $f(R)$ or the non-minimal scalar, the action is equivalent by field redefinition to GR with minimal coupling, so necessarily a second law still holds \cite{Jacobson:1995uq,ford2001classical}. However for the next simplest case of Lovelock gravity, the second law can be violated when two black holes merge \cite{Jacobson:1993xs,Liko:2007vi,Sarkar:2010xp}, even though it holds in certain perturbative regimes \cite{Wall:2015raa, Sarkar:2013swa,Kolekar:2012tq,Bhattacharyya:2016xfs}.  The likely moral is that such theories cannot be completed into a UV-finite theory of quantum gravity unless the problematic couplings are suppressed to the Planck or string scale \cite{camanho2016causality}, in which case, other competing effects must also be taken into account. See \cite{Chatterjee:2013daa} for a partial analysis, although I do not agree with their claim that entropy can be defined only in equilibrium.}

\subsection{Generalized Second Law}

At the quantum level, the NEC is violated by the Hawking effect.  Instead we wish to derive a generalized second law (GSL), which states that $S_{\mathrm{gen}}$ monotonically increases with time on some horizon.  In the semiclassical regime, the GSL is most interesting when expanding around a classically stationary background, so that $\Delta A/4G\hbar$ and $\Delta S_\mathrm{out}$ are the same order in an $\hbar G$ expansion.  Some early limited proofs of the GSL are reviewed in \cite{10proofs}.

The most general derivation of the GSL for causal horizons depends on the fact that the relative entropy between two states $\rho$ and $\sigma$, defined as
\be\label{relS}
S(\rho\,|\,\sigma) \equiv \mathrm{tr}(\rho \ln \rho) - \mathrm{tr}(\rho \ln \sigma)
\ee
is monotonically decreasing when the states are restricted to a subalgebra \cite{Araki:1976zv}, e.g. when information is lost between two cuts of a causal horizon $C^+$.  In the semiclassical regime described above, it turns out that for all cuts, $S(\rho\,|\,\rho_{HH}) = -S_\mathrm{gen}$ (up to an additive constant) \cite{wall2011}, where $\rho_{HH}$ is the Hartle-Hawking state  defined in Section \ref{sec:HH}.  The proof uses \eqref{K} together with a linearization of the Raychaudhuri equation \eqref{eq:raychau} to show that for each horizon cut $v_*(y)$ and state $\rho$,
\be
\mathrm{tr}(\rho \ln \rho_{HH}) = -\frac{\langle A \rangle}{4G\hbar} + c
\ee
where the constant $c$ is independent of $v_*(y)$.  From this one can derive that the GSL holds in a differential sense as the cut of $C^+$ is pushed to the future \cite{wall2011}:
\be\label{GSL}
\frac{\delta}{\delta v_*(y)} S_\mathrm{gen}(v_*) \ge 0.
\ee
This inequality holds in \emph{every} semiclassical state of the matter fields.  No time asymmetric assumption is needed besides the fact that we are considering $C^+$ rather than $C^-$.

On the other hand, for a future trapping horizon $T^+$, it turns out that $S_\mathrm{gen}$ does not always increase \cite{wall2011b}.  But that is because $T^+$ is defined as a surface foliated by leaves whose area $A$ is stationary in a null direction.  In the quantum regime, it is more natural to consider a future \emph{Q-screen} $Q^+$, defined analogously as a (not necessarily null) surface foliated by leaves whose generalized entropy $S_\mathrm{gen}$ is stationary in the null direction\footnote{Surfaces with $\delta S_\mathrm{gen}/{\delta v_*} < 0$ are called \emph{quantum trapped}; given certain assumptions we can use the GSL (in place of the NEC) to prove a quantum singularity theorem given the existence of such surfaces \cite{wall2013generalized}.}  It is then possible to prove that the leaves of a Q-screen obey a GSL (i.e. $S_\mathrm{gen}$ montonically increases) \cite{Bousso:2015eda}, assuming the Quantum Focusing Condition (QFC) described in the next section.

\subsection{Quantum Focusing}

The QFC \cite{bousso2016quantum} states that on a cut of \emph{any} null surface (not necessarily a horizon), the second functional derivative\footnote{These functional derivatives are densitized, i.e. they represent the increase per area element $\delta A$} of the generalized entropy is negative:
\be \label{eq:qfc}
\frac{\delta}{\delta v_*(y)} \frac{\delta}{\delta v_*(y')} S_\mathrm{gen}(v_*) \le 0.
\ee
This generalizes the classical focusing inequality\eqref{Cfocus} to quantum situations.  The QFC is stronger than the GSL on either $C^+$ or $T^+$, and was motivated by the desire to prove a quantum version \cite{Strominger:2003br} of the generalized covariant entropy bound \cite{GCEB}.\footnote{This is one of several proposed bounds on the maximum entropy that can be contained in a given region (e.g. \cite{bekenstein1981universal,Bousso:1999xy,GCEB}).  However, the divergence of entanglement entropy in QFT makes it difficult to give a precise definition of these ``entropy bounds''.  When the effects of Hawking radiation are fully taken into account, no substantive entropy bound is needed for the validity of the GSL \cite{unruh1982acceleration,Marolf:2003wu,Marolf:2003sq,10proofs,wall2011}.  The modern approach to the subject, pioneered by Casini and collaborators \cite{Casini:2008cr,Blanco:2013lea,GCEBq,Bousso:2014uxa}, is to recast the entropy bounds into a form where they automatically hold in any QFT, e.g. because of the positivity or monotonicity of relative entropy \eqref{relS}.}  Presumably it originates from some deep fact about the nature of quantum gravity microstates on a null surface.

For $y \ne y'$, the QFC follows from strong subadditivity (another form of montonicity of relative entropy), so the most interesting part of \label{ref:qfc} is the local piece proportional to a delta function $\delta(y - y')$.  On a nearly stationary null surface perturbed by quantum fields, this reduces to a lower bound on the stress tensor called the quantum null energy condition (QNEC) \cite{bousso2016quantum},
\be
\langle T_{vv} \rangle \ge \frac{\hbar}{2\pi}S^{\prime\prime},
\ee 
where $S^{\prime\prime}$ is the local piece of the second null derivative of the entropy, evaluated on either side of the null surface.  This inequality applies to arbitrary excited states, and has been proven in many field theories,\footnote{The QNEC (and hence the QFC in this regime) has been shown for free or superrenormalizable scalars \cite{Bousso:2015wca}, for holographic theories \cite{Koeller:2015qmn}, and for general $D > 2$ field theories that flow to an interacting conformal fixed point in the ultraviolet \cite{Balakrishnan:2017bjg}---oddly, the QNEC appears to be saturated in this last case \cite{Leichenauer:2018obf,BCFLSunpub}---and there are plausible indications that it holds in $d = 2$ as well \cite{wall2011b,Wall:2017blw}.  There may even be a quantum \emph{dominant} energy condition (QDEC) \cite{Wall:2017blw}.} confirming the QFC in these cases.

\section{Holographic Black Holes} \label{sec:holographic}

After arguing for the validity of the holographic principle, we will describe some profound implications for the information inside of black holes.

\subsection{The Information Puzzle}

What is the ultimate reason why black holes obey a second law?  The majority opinion 
(reviewed in \cite{Polchinski:2016hrw})\footnote{But see \cite{Unruh:2017uaw} for a dissenting review giving the arguments for information loss.}) is that information is not actually lost inside of black holes; it only gets scrambled somehow into the near-horizon degrees of freedom, and in principle the information is still accessible from the outside, if we have access to the full quantum gravity microstates discussed in Sec. \ref{sec:induced}.

\begin{figure}[ht]
\centering
\includegraphics[width=.4\textwidth]{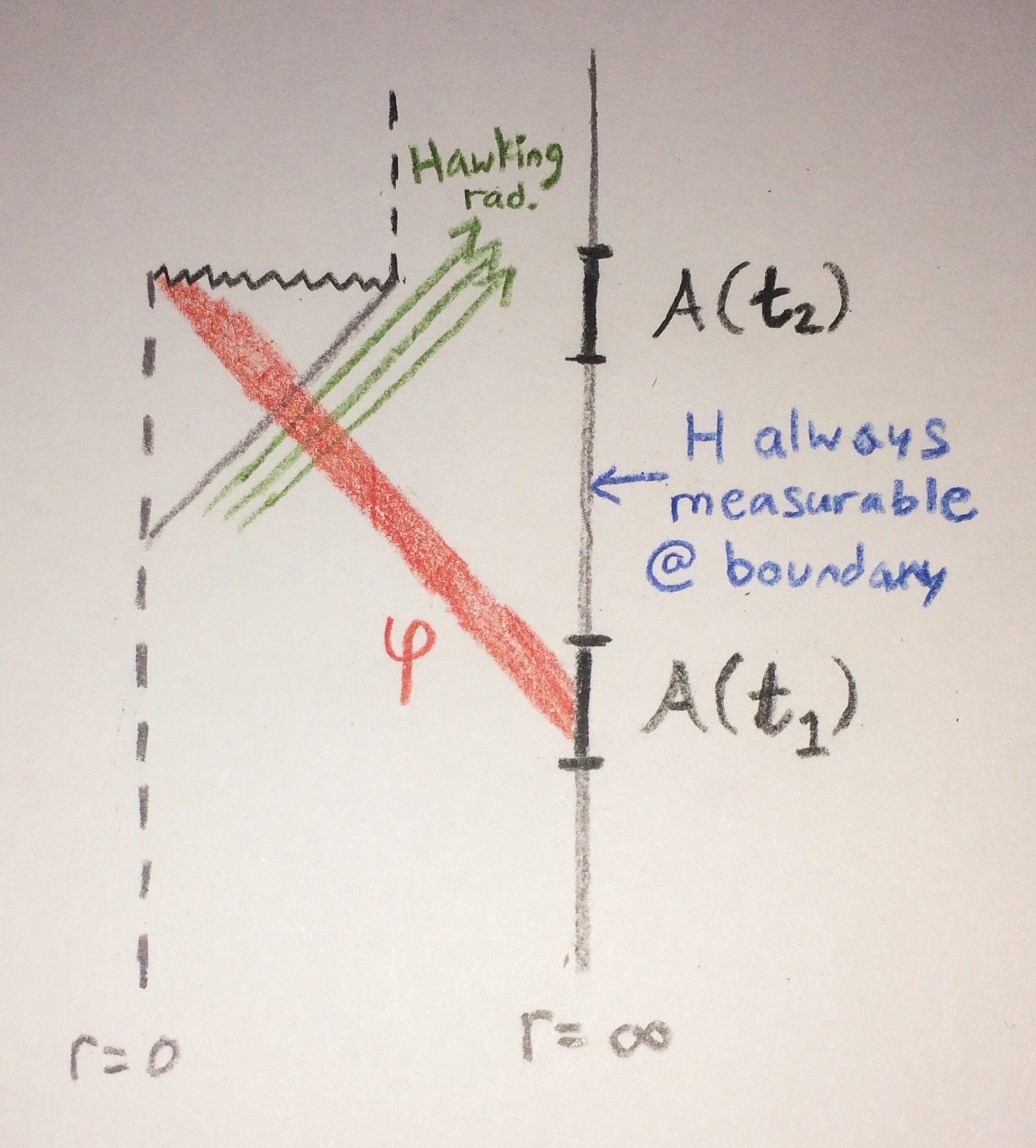}
\caption{\small A black hole is formed from collapse by exciting a scalar field $\varphi$ using operators available in the boundary algebra $\mathcal{A}(t_1)$ at an early time $t_1$.  But because the Hamiltonian $H$ is measurable using the gravitational field at infinity, the details of the collapse are still encoded in the boundary algebra at any later time $t_2 > t_1$, even before the black hole totally evaporates into Hawking radiation!}\label{fig:holo}
\end{figure}

The following argument for information preservation is based on an argument by Marolf \cite{Marolf:2008mf,Marolf:2008mg}: Suppose we have a gravitational theory embedded in a spacetime with a timelike boundary (e.g. an asymptotically AdS spacetime) with suitable reflecting boundary conditions.  (There exist more subtle versions of this argument that apply to asymptotically flat spacetimes.)  Let us define $\mathcal{A}(t)$ as the algebra of observables accessible at the boundary in a small interval around time $t$ (Fig. \ref{fig:holo}).

We assume that the algebras satisfy the following axioms:
\begin{enumerate}
\item  $\mathcal{A}(t)$ is closed under addition, multiplication and reasonable limits (such as exponentiation).  [QM]
\item
$\mathcal{A}(t)$ includes the Hamiltonian $H$, since in diffeomorphism-invariant theories such as GR, the energy can be measured from the ADM mass \cite{Arnowitt:1962hi} of the gravitational fields at infinity.  [GR]
\item
This Hamiltonian $H$ generates time translations by acting on the Hilbert space of the full system in the usual QM way. \label{Heisenberg} [QM]\,\footnote{To see that Assumption \ref{Heisenberg} (by itself) is compatible with information loss, note that it holds for QFT on a fixed spacetime with a Killing horizon, where $H$ is the Killing energy of the quantum fields.}
\item
$\mathcal{A}(t)$ includes nontrivial field operators (e.g. a scalar field $\varphi$ smeared on the boundary near $t$) whose excitations can propagate into the bulk, and form a black hole from collapse.  [AdS field theory]
\end{enumerate}
Now there will be many ways of forming a black hole from boundary operators. But any such operator can also be written in the algebra of observables $\mathcal{A}(t_1)$ at any later time $t_2 > t_1$ using the Heisenberg formula:
\be
\varphi(t_1) = e^{i(t_2 - t_1)H} \varphi(t_2) e^{-i(t_2 - t_1)H}.
\ee
Hence the information in $\varphi$ used to form the black hole is still accessible at time $t_2$ on the boundary for a sufficiently precise experiment.  We conclude that the information accessible on the boundary must evolve unitarily.\footnote{This argument allows for the possibility that there might be some information behind horizons which is never located on the boundary \cite{Marolf:2012xe}.}

Although we did not assume the holographic principle \emph{a priori}, this conclusion is in agreement with AdS/CFT\footnote{For a review see \cite{Hubeny:2014bla}.}, a duality (passing many highly nontrivial checks) which relates string/M-theory on backgrounds with negative cosmological constant (the AdS) to conformal field theories living on the timelike boundary (the CFT).\footnote{Although AdS/CFT applies only when the asymptotic structure of spacetime has a negative cosmological constant $\Lambda$, black hole thermodynamics behaves similarly regardless of the sign of $\Lambda$.  It is therefore plausible that many of these holographic conclusions will continue to hold.}  In the limit where the boundary theory has a large number of species $N$ and strong coupling $\lambda$, the bulk theory becomes classical GR.  But the CFT has unitary time evolution just like any other QFT.

On the other hand, unless information is lost, it seems difficult or impossible to avoid large violations of semiclassical physics in regions far from the singularity \cite{Mathur:2009hf,Almheiri:2012rt,Almheiri:2013hfa,Braunstein:2009my}.  In particular, it is hard to avoid the conclusion that either (i) sufficiently old black holes have no smooth interior (the ``firewall''), or (ii) the quantum modes inside the horizon are reconstructed from the boundary in a nonlinear way (violating the usual measurement rules of QM).  See Harlow \cite{harlow} for a helpful review of some of these puzzles.

\subsection{Holographic Entropy Formula}\label{HEF}

Let us proceed on the assumption that the holographic principle is true and the boundary information is preserved.  In this case, the increasing generalized entropy $S_\mathrm{gen}$ of a dynamically evolving black hole must really be a \emph{coarse-grained} entropy, i.e. an entropy obtained by forgetting some information associated with the microscopic degrees of freedom (just as in the thermodynamics of an ordinary system).  The fine-grained entropy (i.e. $S_{bdy} \equiv -\mathrm{tr}\left(\rho \ln \rho \right)$ of the boundary density matrix $\rho$) would then be constant with time.

How do we measure the fine-grained entropy of a black hole?  In order to ensure that the black hole is not in a pure state, it is convenient to consider the case of a two-sided black hole (which might be dynamically evolving), which in AdS/CFT corresponds to an entangled state of the two boundary CFT's.  See Fig. \ref{fig:holo}. Then the \emph{Hubeny-Rangamani-Takayanagi (HRT) formula} tells us that the leading order boundary entropy (on either side) is given by the area of an ``extremal'' surface $X$ in the bulk \cite{Hubeny:2007xt,Headrick:2007km,Ryu:2006bv}:
\be\label{HRT}
S_{bdy} = \frac{A[X]}{4G\hbar}
\ee
where $X$ is a compact $D-2$ dimensional surface that divides the two boundaries of the wormhole, such that $\theta_{u} = \theta_{v} = 0$, i.e. it is marginally trapped in both the past and the future directions (if there happen to be multiple such surfaces, we use the one that minimizes the entropy).\footnote{This is a special case of a more general principle that allows one to calculate boundary entanglement entropies, using extremal surfaces anchored to boundary regions. This formula has been derived by a path integral argument \cite{lewkowycz}, using a clever analytic continuation to extend the Gibbons-Hawking derivation \cite{gibbons1977action} of black hole entropy to surfaces without a Killing boost symmetry.}

Quantum \cite{Barrella:2013wja,Faulkner:2013ana,Engelhardt:2014gca,Dong:2017xht} or higher curvature \cite{solodukhin2008entanglement,fursaev2013distributional,camps2014generalized,dong} bulk effects are dealt with by replacing the area with a generalized entropy functional, exactly as in sections \ref{sec:BHent} or \ref{sec:2nd}.  Including the leading bulk loop corrections, this tells us that 
\be\label{FLM}
S_{bdy} = \langle A_\mathrm{gen}[X]\rangle + S_\mathrm{bulk}[X],
\ee
where $S_\mathrm{bulk}[X]$ is the entropy in the spacetime region on one side of $X$ (known as the \emph{entanglement wedge}) and $\langle A_\mathrm{gen}[X]\rangle$ is the surface term (whose leading order piece is $A/4G\hbar$).

By taking the first order variation $\rho + \delta \rho$ of \eqref{FLM}, it is also possible to derive \cite{Jafferis:2015del,Dong:2016eik} a corresponding linear operator equation relating the bulk and boundary modular Hamiltonians $K^{(\rho)} = -\ln \rho$:
\be\label{modHam}
K^{(\rho)}_{bdy} = A_\mathrm{gen}[X] + K^{(\rho)}_\mathrm{bulk}[X].
\ee 
This is a remarkable formula since in general both $K$'s are highly nonlocal, and $\rho$ could be any QFT state in the semiclassical regime!  This equivalence suggests ways to reconstruct data inside the entanglement wedge by using the ``modular flow'', i.e. the evolution defined using $K$ as a Hamiltonian \cite{Jafferis:2015del,Faulkner:2017vdd,Almheiri:2017fbd}.  (Previous reconstruction techniques only allowed one to reconstruct data outside the event horizon \cite{HKLL}.)

Combining \eqref{modHam} and \eqref{FLM}, we see that the relative entropy \eqref{relS} of the boundary also agrees with the relative entropy in the entanglement wedge:
\be
S_\mathrm{bdy}(\rho \,|\, \sigma) = \langle K^{(\sigma)} \rangle_\rho - S(\rho) = 
S_\mathrm{bulk}(\rho \,|\, \sigma)[X].
\ee
This formula can be used to derive \emph{entanglement wedge reconstruction}, the idea that---in a suitable ``code subspace'' of states in which semiclassical physics is valid \cite{Almheiri:2014lwa,Dong:2016eik}---the data on just one of the two boundary CFT's fixes all the data in the entanglement wedge on its side of $X$ (and vice versa)
\cite{Czech:2012bh,Wall:2012uf,Headrick:2014cta,Pastawski:2015qua,Dong:2016eik}. 
 
\subsection{Thermalization and Scrambling}

This leaves the question of how to interpret the horizon area of a nonstationary black hole.  The fluid-gravity correspondence relates the long-wavelength dynamics on a horizon to hydrodynamics on the CFT side \cite{Hubeny:2011hd}.  However, if we wish to provide a stat-mech interpretation of the growing areas of $C^+$ or $T^+$, then we need to identify some specific coarse-graining procedure that gives rise to the entropy increase.

So far this has been done only for spacelike trapping horizons $T^+$ (satisfying mild additional assumptions), where it can be shown using HRT \eqref{HRT} and the NEC that the area of each marginally trapped surfaces $\mu$ in $T^+$ is proportional to the maximum boundary entropy $S_\text{bdy}$ that is compatible with the classical data outside of $\mu$ \cite{Engelhardt:2017aux,Engelhardt4th}:
\be
S_\text{coarse} \equiv \text{max}(S_\text{bdy} \,:\, \text{data outside of }\mu) = \frac{A[\mu]}{4G\hbar}.
\ee
This naturally explains the area-increase theorem in the case of spacelike $T^+$'s, since as we are maximizing the entropy subject to fewer constraints as we move outwards.  \cite{Engelhardt:2017aux,Engelhardt4th} argue that this is dual to a coarse-grained second law of thermodynamics on the boundary CFT.  (It is not yet known how to interpret the area of the causal horizon $C^+$ as a coarse-grained entropy \cite{Hubeny:2012wa,Kelly:2013aja}, and several conjectures \cite{Freivogel:2013zta,Kelly:2013aja} have already been falsified \cite{Engelhardt:2017wgc}.)

Even after the boundary CFT has reached its final coarse-grained entropy (thermalization), there is a somewhat longer timescale required for the information in a single degree of freedom to acquire a large commutator with every other degree of freedom (scrambling) \cite{Hayden:2007cs,Sekino:2008he}.\footnote{The ratio between the two timescales is $t_S / t_T \sim \ln(R/L_{\text{planck}})$ where $R$ is the black hole radius.}  The scrambling time can be calculated holographically from the gravitational interaction of shock waves propagating near the black hole horizon \cite{Shenker:2013pqa}.  Scrambling physics can be used to facilitate quantum teleportation between two entangled CFT's, which turns out to be holographically dual to a traversable wormhole \cite{Gao:2016bin,Maldacena:2017axo,Susskind:2017nto}.  It also plays an important role in recent speculation involving the quantum complexity (number of gates in a quantum circuit) of black hole states \cite{Stanford:2014jda,Brown:2015bva} and the firewall paradox \cite{Almheiri:2012rt,Almheiri:2013hfa}.  

Hopefully all these vexing issues will become more clear once we have a good understanding of how to reinterpret the geometry of spacetime in terms of the quantum information flowing through various surfaces.  From this perspective, black hole horizons are just an illuminating special case of principles that should be valid everywhere in the universe.

\subsubsection*{Acknowledgements}
{\small
This work was supported by the Stanford Institute for Theoretical Physics, the Simons Foundation (``It from Qubit''), and AFOSR grant number FA9550-16-1-0082.  I am grateful to Geoff Penington (for typing assistance), Patrick Hayden (for letting me have Geoff), Nathan Benjamin (a test reader), Jon Sorce (for explaining his paper to me), and my former Ph.D. advisor Ted Jacobson (for teaching me the subject).  
}

\bibliographystyle{utcaps}
\bibliography{all}

\end{document}